\documentclass[aps,prl,twocolumn,groupedaddress,amssymb,amsmath]{revtex4}
\usepackage{amssymb}

\usepackage{graphicx}
\usepackage{epsf}
\usepackage{marvosym}
\usepackage{latexsym}

\begin{document}

\title{Possible evidence for two-gap superconductivity in $Cu_xTiSe_2$}
\author{ Moran Zaberchik$^{1}$, Khanan Chashka$^{1}$, Larisa Patlgan$^{1}$, Ariel Maniv$^{2}$, Chris Baines$^{3}$, Philip King$^{4}$ and Amit Kanigel$^{1}$}
\affiliation{$^{1}$Physics Department, Technion-Israel Institute of Technology, Haifa 32000, Israel}
\affiliation{$^{2}$NRCN, P.O.Box 9001, Be'er Sheva, Israel}
\affiliation{$^{3}$Paul Scherrer Institute, CH 5232 Villigen PSI, Switzerland}
\affiliation{$^{4}$Rutherford Appleton Laboratory, Chilton Didcot, Oxforshire OX11 0QX, United Kingdom}

\begin{abstract}
We report a comprehensive TF-$\mu$SR study of $\rm{Cu_xTiSe_2}$. The magnetic penetration depth was found to saturate at low temperature as expected in an s-wave SC. 
As $x$ is increased we find that the superfluid density increases and the size of the superconducting gap, calculated from the temperature dependence of the superfluid density, is approaching the BCS value. However, for low values of $x$, the gap is smaller than the weak-coupling BCS prediction suggesting that two superconducting gaps are present in the sample.

\end{abstract}
\maketitle

 The transition-metal dichalcogenides (TMDCs) exhibit a variety of interesting physical phenomena \cite{review}. Many of the TMDCs have a charge-denisty wave (CDW) modulation in the ground state and many others show superconductivity (SC) in the ground state. There are few examples of systems having both CDW and SC, which raises interesting questions regarding the way the Fermi surface accommodates these, apparently, competing orders. The interplay between CDW and SC is re-gaining attention, mainly in the context of the newly discovered pnictide high temperature superconductors \cite{FeAs_review}. 
 
 Recently, superconductivity was discovered in Cu intercalated TiSe$_2$ \cite{Cava}. The addition of Cu gradually suppresses the CDW transition temperature and above 4\% Cu a superconducting phase emerges. T$_c$ increases with the Cu concentration up to 4.15K for x=0.08, and decreases for higher concentrations.  This compound provides an opportunity to study in detail the interplay between CDW and SC by simply changing x. 
  
  We report a detailed study of the  SC state in  $\rm{Cu_xTiSe_2}$, using Muon Spin Rotation ($\mu$SR). $\mu$SR allows us to measure the magnetic penetration depth, $\lambda$, as a function of the Cu doping, and the temperature dependence of $\lambda$, which is  intimately related to the SC gap \cite{Sonier}.

Pressed powder samples of $\rm{Cu_xTiSe_2}$ with different Cu amounts, ranging from x=0.02 to x=0.081, were prepared as described in Ref.~\cite{Cava}. Copper concentration was calculated using X-ray diffraction (XRD) and the known c-axis calibration~\cite{Cava}. XRD patterns show no signs of foreign phases. 

\begin{figure}
  \begin{center}
 \includegraphics[width=9.5cm]{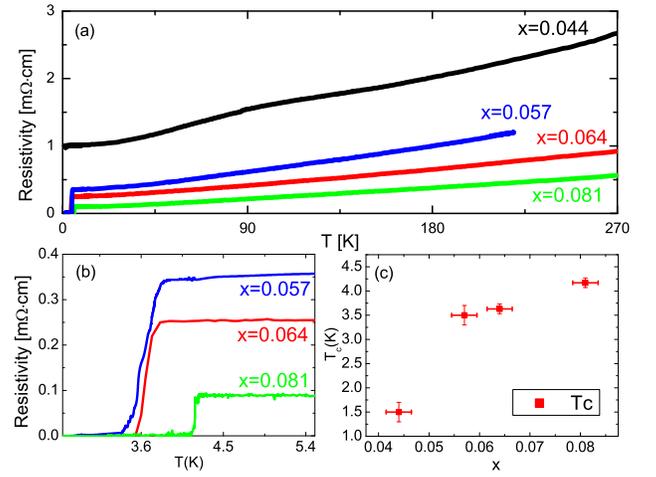}
 \end{center}
  \caption{(a) The temperature dependence of the resistivity for 4 samples with various Cu concentration.  (b) Resistivity at low temperatures. (c) $\rm{T_c}$ Vs. x.}
 \label{RVsT}
\end{figure}

The resistivity as a function of the temperature for all the samples is presented in Fig.~\ref{RVsT}a. All the samples show metallic behavior at high temperatures.  The x=0.044 sample has, in addition, a broad hump at around 100K. This is related to the formation of the CDW in the sample. Above x=0.044 this broad hump cannot be observed anymore, consistent with the CDW phase ending at around 4\% Cu.  

A zoom-in on the low temperature part for three of the samples is shown in Figure~\ref{RVsT}b.  In Figure~\ref{RVsT}c $\rm{T_c}$ vs. the Cu doping can be seen. The T$_c$ of the x=0.044 sample was measured using $\mu$SR, since we cannot measure resistivity below 1.8K. So far, the preparation of $\rm{Cu_xTiSe_2}$ was reported by two groups \cite{Cava, TiSe2CuX_single_crystals}; they find  very similar phase diagrams. In Ref.~\cite{Cava} superconductivity emerges just above x=0.04. Both groups report x=0.11 to be the solubility limit of Cu. 
We find that the x=0.044 sample is superconducting with a transition temperature of about 1.5K; samples with lower Cu concentration are not superconducting down to 50mK as revealed by $\mu$SR.
The maximal $\rm{T_c}$ is 4.17K, for x$\approx$0.08, similar to the maximal $\rm{T_c}$ reported previously~\cite{Cava}.

\begin{figure}
  \begin{center}
  \includegraphics[width=8cm]{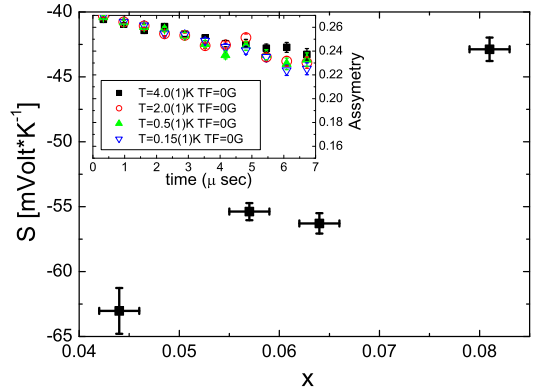}
\end{center}
  \caption{The Cu concentration dependence of the thermoelectric-power (TEP) at room temperature. Inset: ZF-$\mu$SR asymmetry for the x=0.08 sample at various temperatures.  }
 \label{SGB}
\end{figure}

In Fig.~\ref{SGB} we show thermoelectric-power (TEP) data taken at room temperature for the samples. The results indicate an increase in the carrier density as Cu is added. The Seebeck coefficient is negative for all the samples at all temperatures as expected  for negative charge carriers (electrons). This is in agreement with previous results~\cite{Cava,TiSe2CuX_single_crystals}.

Zero-Field (ZF) $\mu$SR measurements were done at the Rutherford Appleton Laboratory, using a dilution refrigerator (DR) and a flow cryostat. ZF $\mu$SR is a very sensitive local probe of magnetic field. The ZF $\mu$SR results for the x=0.081 sample are shown in the inset of Fig. \ref{SGB}.  We see a very weak gaussian relaxation for all temperatures down to 150mK, with no sign of magnetism. 
This suggests that probably each Cu ion contributes one electron to the sample, and that the Cu ions are in the Cu$^{+1}$ configuration. Cu$^{++}$ is magnetic and thus is ruled out.

Transverse-Field (TF) $\mu$SR measurements were done at the Laboratory for Muon Spin Spectroscopy (LMU) in the Paul Scherrer Institute (PSI), using the General Purpose Surface-Muon (GPS) and the Low Temperature Facility (LTF) instruments.

For the $\mu$SR experiments the samples were mounted on silver backing plates. The samples were cooled in a magnetic field of 1000G from above T$_c$ down to 50mK. The magnetic field direction was perpendicular to the muons' initial polarization direction.  
In addition, we measured the x=0.081 sample at 1.7K in magnetic field ranging from 400G to 1500G. 

In order to analyze the TF-$\mu$SR data we used the following fitting function:
\begin{equation}
\begin{split}
A(t)=A_{sc} \exp \left( -\frac{\sigma^2_{dip}+\sigma^2_{sc}}{2} t^2 \right) \cos(\gamma_{\mu} B_{int} t + \phi) + \\ A_{Ag} \exp(-\frac{\sigma^2_{Ag}}{2}t^2)\cos(\gamma_{\mu} B_{ext} t).
\end{split}\end{equation}
The first term is the contribution of the sample, while the second term is the contribution of the silver holder. 
$A_{sc}$ and $A_{Ag}$ are the initial asymmetries, $B_{int}$ and $B_{ext}$ are the internal field inside the SC and the external applied field, respectively.  $\gamma_{\mu}= 2 \pi \times 135.5342$ MHz/T is the  gyromagnetic ratio of the muon, and $\phi$ is the initial phase.    
There are two contributions to the relaxation of the muons in the sample, $\sigma_{sc}$, the origin of which is the inhomogeneity of the field due to the formation of the vortex-lattice in the sample, and $\sigma_{dip}$, the contribution to the dipolar relaxation due to the nuclear moments. $\sigma_{Ag}$ is the very slow relaxation of the muons in silver. 

The data for each sample over the entire temperature range are fitted globally. The initial asymmetries, the initial phase, $\sigma_{dip}$ and the external field are temperature-independent.  $\sigma_{Ag}$ was measured independently. This allows us to reduce the number of free parameters and increase the accuracy of the fitting procedure. $\sigma_{sc}$  is  proportional $1/\lambda^2$, which,  through the London relation, is proportional to the superfluid density of the sample \cite{musr_book}. For anisotropic materials, with an anisotropy ratio larger than five, it was shown that the $\mu$SR line width is proportional to $1/\lambda_{ab}^{2}$, where $\lambda_{ab}$ is the in-plane penetration depth \cite{Barford}. As $\rm{Cu_xTiSe_2}$ is very anisotropic \cite{AniCava}, all our measurements are sensitive only to the in-plane part of the penetration depth. 
 
\begin{figure}
  \includegraphics[width=8cm]{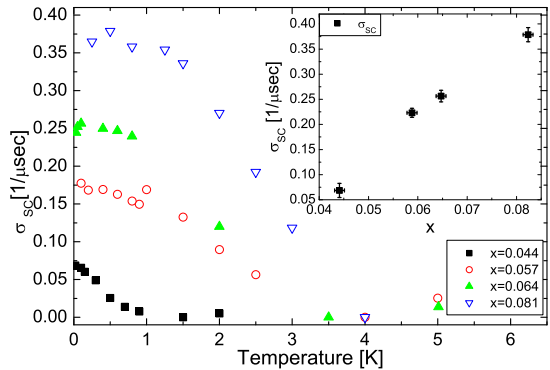}
  \caption{ The temperature dependence of the relaxation due to the flux-lattice formation for various samples with different copper amounts. Inset: $\sigma_{sc}$ at the lowest temperature Vs. x.}
 \label{RlxVsTandX}
\end{figure}

The temperature dependence of $\sigma_{sc}$ for all the samples is presented in Fig.~\ref{RlxVsTandX}. It can be seen that the relaxation increases as the temperature decreases, and that it saturates at low temperatures. In an s-wave SC in which there is a fully gapped Fermi surface, a thermally activated behavior is expected. This leads to an almost constant superfluid density at low temperatures. In contrast, nodes in the gap lead to a linear temperature dependence at low temperatures.  Our results are consistent with an s-wave order parameter in $\rm{Cu_xTiSe_2}$, in agreement with previous work \cite{Taillefer}. 

In the inset of Fig. \ref{RlxVsTandX}, we show the low temperature value of the muon relaxation versus the Cu content x. We find that it increases linearly when Cu is added to the system.  In a BCS SC T$_c$ is not related to the superfluid density, in general. Here, we believe that because the changes induced by the Cu on the structure of the sample are very small, it will have a negligible effect on the electron-phonon interaction and on the pairing potential. Cu increases the amount of free carriers which will raise the Fermi-energy and increase the density of states at the Fermi energy and will increase T$_c$. Each Cu adds one mobile electron to the system, and that electron contributes to the superfluid density below T$c$.

In general, extracting the penetration depth from the $\mu$SR line width, $\sigma$, is not a simple task \cite{Landau}. For the case of a perfect vortex-lattice in a single-gap s-wave SC, there is a very good approximation by Brandt \cite{Brandt03}, based on a numerical solution of the Ginzburg-Landau equations. This approximation works for  $\kappa \geq 5$ and for a wide range of fields, $0.25/\kappa ^{1.3}  \le B/B_{c2} \le 1$ where $\kappa$ is the Ginzburg-Landau parameter:

\begin{equation}
\label{brandt}
\begin{split}
\sigma_{sc}[\mu s^{-1}]=  4.83 \cdot 10^4  \times \left( {1 - b} \right) \times \\ \left[ {1 + 1.21\left( {1 - \sqrt b } \right)^3 } \right]\lambda ^{ - 2} [ {nm} ]
\end{split}\end{equation}
where $b=B/B_{c2}$. 

\begin{figure}
\begin{center}
  \includegraphics[width=8cm]{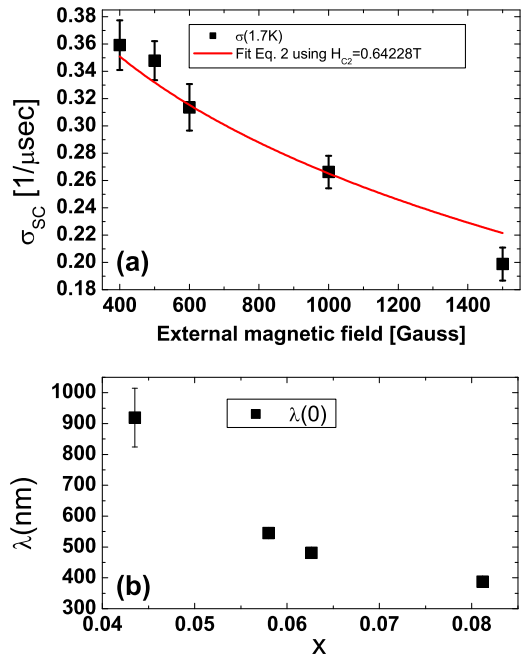}
\end{center}
  \caption{(a) The dependance of the relaxation on the external magnetic field for the x=0.081 sample. (b) $\lambda$ Vs. the copper ratio (x).}
 \label{LVsHandAsyVstime}
\end{figure}

A possible self-consitency check is to verify that the penetration depth calculated using Eq. \ref{brandt} is field-independent. The penetration depth measured by $\mu$SR is calculated by measuring the length-scale associated with the decay of the super-currents around the vortex-core. This length-scale is expected to be field-independent for a single s-wave gap SC \cite{Landau}. 
In Fig. \ref{LVsHandAsyVstime}a, we show the field dependence of $\sigma_{sc}$ for the $x=0.081$ sample, taken at T=$1.7K$. The solid line is a fit to the data using Eq. \ref{brandt}, and H$_{c2}(T=1.7K)=0.64T$ as determined from magneto-resitance data, $\rho$(H), measured using the same sample. We get $\lambda = 441 \pm 5$ nm,  and as one can see, the agreement is reasonable, but not perfect. A possible explanation for this very weak field dependence of $\lambda$ is that this sample is not a perfect single gap s-wave SC.

We use the Gorkov approximation for the temperature dependence of H$_{c2}$ \cite{HC2_V_Temperature_Gorkov} to extrapolate H$_{c2}$(T) to low temperatures. We obtain H$_{c2}$(0)=0.85T for the $x=0.081$ sample; we can then estimate the zero-temperature coherence-length $\xi(0)=\sqrt{\Phi_0/ 2 \pi H_{c2}}$ to be 180\AA. The fact that $\lambda$ does not depend on the field and that we get $\kappa>5$ indicates that Eq. \ref{brandt} can be used for analyzing our $\mu$SR data, at least for the $x=0.081$ sample.
Next, we used Eq. \ref{brandt} to calculate $\lambda$(T) using the data presented in Fig. \ref{RlxVsTandX}a for three of our samples. It is essential to take into account the temperature dependence of H$_{c2}$ when calculating $\lambda$. For each sample we measured H$_{c2}$ at high temperature and used the Gorkov approximation \cite{HC2_V_Temperature_Gorkov} to extrapolate H$_{c2}$ to lower temperatures. We found that H$_{c2}$ depends rather weakly on the doping. The penetration depth at base temperature for all the samples is shown in Fig. \ref{LVsHandAsyVstime}b.

\begin{figure}
\begin{center}
  \includegraphics[width=8cm]{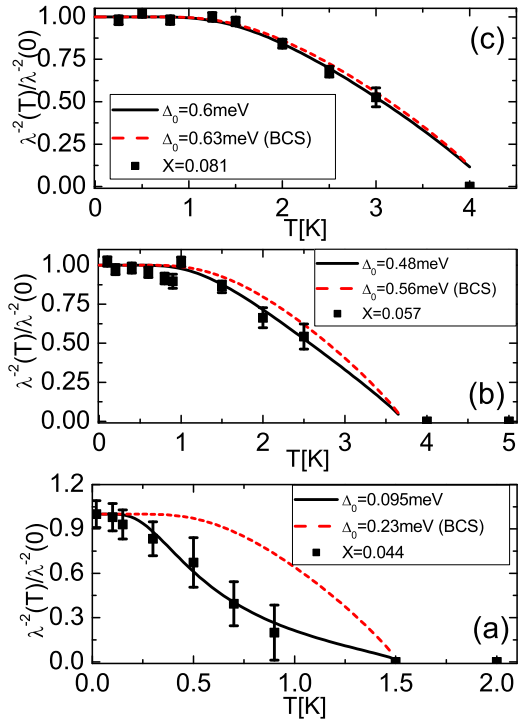}
\end{center}
  \caption{Temperature dependence of $\lambda(T)^{-2}$. The black solid lines are best fits to the data using Eq. \ref{SC_gap_fit}. The dashed red lines are the BCS prediction. Figure (a) for x=0.044(1) sample, Figure (b) for x=0.057(1) sample, and Figure (c) for x=0.081(1) sample.}
 \label{SC_gaps}
\end{figure}

The temperature dependence of $\lambda$, as already mentioned, is set by the SC gap size and momentum dependence.  $\lambda(T)^{-2}$ for all the samples saturates at low temperature indicating the s-wave nature of the gap.  
The gap size can be extracted from the data using this relation\cite{Tinkham}:
\begin{equation}\label{SC_gap_fit}
{{\lambda ^{ - 2} \left( T \right)} \over {\lambda ^{ - 2} \left( 0 \right)}} = 1 - 2\int\limits_{\Delta \left( T \right)}^\infty  {\left( { - {{\partial f} \over {\partial E}}} \right){E \over {\sqrt {E^2  - \left( {\Delta \left( T \right)} \right)^2 } }}dE}
\end{equation}
where $f$ is the Fermi function, $\Delta(T)$ is the temperature-dependent superconducting gap and the integration is over energy measured from the chemical potential.   We used $\Delta \left( {\rm{T}} \right) = \Delta \left( 0 \right)\tanh \left\{ {1.82\left[ {1.018\left( {{{{\rm{T}}_{\rm{C}} } \over {\rm{T}}}} \right) - 1} \right]^{0.51} } \right\}
$ to describe the temperature dependence of the gap \cite{khasanov_gap_Vs_Temperature}. The data and the fitting curves are shown in Fig. \ref{SC_gaps}. The black solids line represent the best fit, and the red dashed lines represent the expected temperature dependence for the gap size predicted by the weak-coupling BCS theory, $\Delta(0)$=1.764k$_{b}$T$_{c}$. The gap sizes we find are: $\Delta(0)$=0.60(2) meV, $\Delta(0)$=0.48(2) meV and $\Delta(0)$=0.095(5) meV for the $x=0.081$, $x=0.057$ and $x=0.044$, respectively.

We systematically get gap values that are lower than the BCS weak-coupling values,
$\Delta \left( 0 \right)$ = 1.764k$_B$. While for the $x=0.081$ sample, $\Delta$/ k$_{B}$T$_{c}$=1.67 $\pm$ 0.09 is in reasonable agreement with the BCS value, for the lower $T_c$ samples we get gap-to-T$_c$ ratios of 1.51 $\pm$0.09 and 0.7 $\pm$ 0.2.
The small gap values tell us that the simple single s-wave gap picture cannot explain our data. 
$\lambda$(T) is controlled by thermal excitation of quasi-particles across the SC gap. In multi-gap systems or in systems with an anisotropic gap $\lambda$(T) is a measure of the smallest gap in the system.  
Similar results, namely small $\Delta(0)$/T$_c$,  have been found in MgB$_2$ \cite{MgB2 two gaps} and in NbSe$_2$ \cite{NbSe2 two gaps}, too.  MgB$_2$ has two superconducting gaps as was seen, for example, using ARPES \cite{Souma_MgB2}.  
NbSe$_2$, has a multi-sheet Fermi-surface \cite{NbSe2 FS}, with two electron-like cylindical Fermi-surface derived from the Nb band, centered around the $\Gamma$ and $K$ points. In addition, there is a holelike  Fermi-sheet derived from the Se band around the $\Gamma$ point.
In this system, which is structurally similar to TiSe$_2$, two SC gaps were clearly identified on the two hole-like Fermi sheets \cite{NbSe2_Science}.

Our case is different: We find an evolution of the $\Delta(0)/T_c$ ratio as the doping level is changed. 
Based on our data, and the available models for analyzing $\mu$SR data, we cannot say if in $\rm{Cu_xTiSe_2}$ there are two s-wave gaps residing on two different Fermi-surface sheets or one anisotropic s-wave gap.

$\rm{Cu_xTiSe_2}$ ARPES data \cite{ARPES_Feng,ARPES_Hasan} clearly show two Fermi sheets, one hole-like pocket centered around the $\Gamma$ point and one electron-like pocket around the L points. The electron-pocket was shown to grow with the Cu concentration, while the hole-pocket seems less sensitive to the amount of Cu. Combined with our result that shows that the zero temperature superfluid density increases with the addition of Cu, this might suggest that there is a small SC gap on the hole pocket the contribution of which to the overall superfluid density is decreasing as more Cu is added to the system. 

An alternative scenario is one in which there is a single anisotropic s-wave gap, probably on the large electron-like pocket, which becomes more isotropic as charge is added to the system and the CDW vanishes. Recently it was suggested, based on ARPES data in NbSe$_2$ \cite{arcs_in_NbSe2}, that at the CDW transition, parts of the Fermi surface which are nested are gapped out and excluded from participation in superconductivity. It is not clear to us how a situation like that  would influence the temperature dependence of the penetration depth, but it might lead to an anisotropic SC gap 

In summary, we have performed TF-$\mu$SR experiments on a set of $\rm{Cu_xTiSe_2}$ samples with different Cu concentrations. The saturation of the muon relaxation at low temperatures indicates that the Fermi surface in this system is fully gapped, but the temperature  dependence of the penetration depth cannot be explained by a single isotropic s-wave gap. The data indicate that there are probably two gaps in this system the contribution to which the superfluid density changes with the Cu concentration. 

We would like to thank ISIS and PSI facilities for their 
kind hospitality. We are grateful to A. Keren and R. Khasanov for 
helpful discussions and to K. Alexander and G.M Reisner for help with the experiment.


\begin{thebibliography}{99}


\bibitem{review} J.A. Wilson,F.J. Di Salvo and S. Mahajan, Advances in Physics \textbf{24}, 117 (1975).

\bibitem{FeAs_review} M.R. Norman, Physics \textbf{1}, 21 (2008).




\bibitem{Cava} E. Morosan, H. W. Zandbergen, B. S. Dennis, J. W. G. Bos, Y. Onose, T. Klimczuk, A. P. Ramirez, N. P. Ong and R. J. Cava, Nature Physics \textbf{2}, 544 (2006).
\bibitem{Sonier} J.E. Sonier, J.H. Brewer and R.F. Kiefl, Rev. Mod. Phys., \textbf{72}, 769 (2000).

\bibitem{TiSe2CuX_single_crystals} G. Wu,  H. X. Yang, L. Zhao, X. G. Luo, T. Wu, G. Y. Wang, and X. H. Chen, Phys. Rev. B \textbf{76}, 024513 (2007).
\bibitem{musr_book}  S.L. Lee, S. H. Kilcoyne and R. Cywinski, Muon Science, Taylor and Francis, New York 1999. 
\bibitem{Barford} W. Barford and J.M.F. Gunn, Physica \textbf{C156}, 515 (1988).
\bibitem{AniCava} E. Morosan, Lu Li, N. P. Ong and R. J. Cava, Phys. Rev. B \textbf{75}, 104505 (2007).
\bibitem{Taillefer}  S. Y. Li, G. Wu, X. H. Chen, and Louis Taillefer, Phys. Rev. Lett. \textbf{99}, 107001 (2007).



\bibitem{Landau} I.L. Landau and H. Keller, Physica \textbf{C458}, 38 (2007).
\bibitem{Brandt03} E. H. Brandt, Phys. Rev. \textbf{B68}, 054506 (2003). 
\bibitem{HC2_V_Temperature_Gorkov} A. A. Abrikosov and L. P. Gor'kov, Zh. Experim. i Teor. Fiz. \textbf{39}, 480 (1960); [English transl.: Soviet Phys.JETP \textbf{12}, 337 (1961)].
\bibitem{Tinkham} M. Tinkham, Introduction to superconductivity, Krieger Publishing Company, Malabar, Florida (1975). 
\bibitem{khasanov_gap_Vs_Temperature} R. Khasanov, P. W. Klamut, A. Shengelaya, Z. Bukowski, I. M. Savi?, C. Baines, and H. Keller, Physical Review B \textbf{78}, 014502 (2008).
\bibitem{MgB2 two gaps} F. Manzano, A. Carrington, N. E. Hussey, S. Lee, A. Yamamoto, and S. Tajima, Physical Rev. Lett. \textbf{88}, 047002 (2002).
\bibitem{NbSe2 two gaps} J. D. Fletcher, A. Carrington, P. Diener, P. Rodire, J. P. Brison, R. Prozorov, T. Olheiser, and R. W. Giannetta, Physical Review Letters \textbf{98}, 057003 (2007).
\bibitem{Souma_MgB2} S. Souma,  Y. Machida, T. Sato, T. Takahashi, H. Matsui, S.-C. Wang, H. Ding, A. Kaminski, J. C. Campuzano, S. Sasaki and K. Kadowaki, Nature \textbf{423}, 65 (2003).
\bibitem{NbSe2 FS} Th. Straub, Th. Finteis, R. Claessen, P. Steiner, S. HŸfner, P. Blaha, C. S. Oglesby, and E. Bucher , Phys. Rev. Lett. \textbf{82}, 4504 (1999).
\bibitem{NbSe2_Science} T. Yokoya, T. Kiss, A. Chainani, S. Shin, M. Nohara, H. Takagi, Science \textbf{294}, 5551 (2001).
\bibitem{ARPES_Feng} J. F. Zhao, H. W. Ou, G. Wu, B. P. Xie, Y. Zhang, D. W. Shen, J. Wei, L. X. Yang, J. K. Dong, M. Arita, H. Namatame, M. Taniguchi, X. H. Chen, and D. L. Feng, Phys. Rev. Lett. \textbf{99}, 146401 (2007).

\bibitem{ARPES_Hasan} D. Qian, D. Hsieh, L. Wray, E. Morosan, N. L. Wang, Y. Xia, R. J. Cava, and M. Z. Hasan , Phys. Rev. Lett. \textbf{98}, 117007 (2007).
\bibitem{arcs_in_NbSe2} S.V. Borisenko,  A. A. Kordyuk, V. B. Zabolotnyy, D. S. Inosov, D. Evtushinsky, B. BŸchner, A. N. Yaresko, A. Varykhalov, R. Follath, W. Eberhardt, L. Patthey, and H. Berger, Phys. Rev. Lett. \textbf{102}, 166402 (2009).





\end{thebibliography}
\end{document}